\documentclass[sigconf]{acmart}

\AtBeginDocument{%
 }

\copyrightyear{2023}
\acmYear{2023}
\setcopyright{rightsretained}
\acmConference[KDD '23]{Proceedings of the 29th ACM SIGKDD Conference on Knowledge Discovery and Data Mining}{August 6--10, 2023}{Long Beach, CA, USA}
\acmBooktitle{Proceedings of the 29th ACM SIGKDD Conference on Knowledge Discovery and Data Mining (KDD '23), August 6--10, 2023, Long Beach, CA, USA}
\acmDOI{10.1145/3580305.3599872}
\acmISBN{979-8-4007-0103-0/23/08}

\usepackage{url}
\usepackage[utf8]{inputenc}
\usepackage{caption}
\usepackage{graphicx}
\usepackage{amsmath}
\usepackage{amsthm}
\usepackage{booktabs}
\usepackage{arydshln}
\usepackage{multirow}

\makeatletter
\gdef\@copyrightpermission{
  \begin{minipage}{0.3\columnwidth}
   \href{https://creativecommons.org/licenses/by/4.0/}{\includegraphics[width=0.90\textwidth]{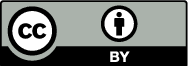}}
  \end{minipage}\hfill
  \begin{minipage}{0.7\columnwidth}
   \href{https://creativecommons.org/licenses/by/4.0/}{This work is licensed under a Creative Commons Attribution International 4.0 License.}
  \end{minipage}
  \vspace{5pt}
}
\makeatother

\begin{document}

\title[Multimodal Indoor Localisation in Parkinson's Disease for Detecting Medication Use: \\ Observational Pilot Study in a Free-Living Setting]{Multimodal Indoor Localisation in Parkinson's Disease for Detecting Medication Use: Observational Pilot Study in a Free-Living Setting}

\author{Ferdian Jovan}
\orcid{0000-0003-4911-540X}
\affiliation{%
  \institution{Faculty of Engineering, University of Bristol}
  \streetaddress{1 Cathedral Square}
  \city{Bristol}
  \country{UK}
  \postcode{BS1 5DD}
}
\email{ferdian.jovan@bristol.ac.uk}

\author{Catherine Morgan}
\affiliation{%
  \institution{Translational Health Sciences, Bristol Medical School}
  \streetaddress{University of Bristol}
  \city{Bristol}
  \country{UK}
  \postcode{BS2 8DZ}
}
\email{catherine.morgan@bristol.ac.uk}

\author{Ryan McConville}
\affiliation{%
  \institution{Faculty of Engineering, University of Bristol}
  \streetaddress{1 Cathedral Square}
  \city{Bristol}
  \country{UK}
  \postcode{BS1 5DD}
}
\email{ryan.mcconville@bristol.ac.uk}

\author{Emma L. Tonkin}
\affiliation{%
  \institution{Faculty of Engineering, University of Bristol}
  \streetaddress{1 Cathedral Square}
  \city{Bristol}
  \country{UK}
  \postcode{BS1 5DD}
}
\email{e.l.tonkin@bristol.ac.uk}

\author{Ian Craddock}
\affiliation{%
  \institution{Faculty of Engineering, University of Bristol}
  \streetaddress{1 Cathedral Square}
  \city{Bristol}
  \country{UK}
  \postcode{BS1 5DD}
}
\email{ian.craddock@bristol.ac.uk}

\author{Alan Whone}
\affiliation{%
  \institution{North Bristol NHS Trust, Southmead Hospital}
  \streetaddress{University of Bristol}
  \city{Bristol}
  \country{UK}
  \postcode{BS10 5NB}
}
\email{alan.whone@bristol.ac.uk}

\renewcommand{\shortauthors}{Ferdian Jovan et al.}

\begin{abstract}
  Parkinson's disease (PD) is a slowly progressive, debilitating neurodegenerative disease which causes motor symptoms including gait dysfunction. Motor fluctuations are alterations between periods with a positive response to levodopa therapy ("on") and periods marked by re-emergency of PD symptoms ("off") as the response to medication wears off. These fluctuations often affect gait speed and they increase in their disabling impact as PD progresses. To improve the effectiveness of current indoor localisation methods, a transformer-based approach utilising dual modalities which provide complementary views of movement, Received Signal Strength Indicator (RSSI) and accelerometer data from wearable devices, is proposed. A sub-objective aims to evaluate whether indoor localisation, including its in-home gait speed features (i.e. the time taken to walk between rooms), could be used to evaluate motor fluctuations by detecting whether the person with PD is taking levodopa medications or withholding them. To properly evaluate our proposed method, we use a free-living dataset where the movements and mobility are greatly varied and unstructured as expected in real-world conditions. 24 participants lived in pairs (consisting of one person with PD, one control) for five days in a smart home with various sensors. Our evaluation on the resulting dataset demonstrates that our proposed network outperforms other methods for indoor localisation. The sub-objective evaluation shows that precise room-level localisation predictions, transformed into in-home gait speed features, produce accurate predictions on whether the PD participant is taking or withholding their medications.
\end{abstract}



\maketitle

\section{Introduction}
\label{sec:introduction}

Parkinson’s disease (PD) is a debilitating neurodegenerative disease affecting around 6 million people worldwide. It is characterised by a variety of movement-related (motor) symptoms, such as slowness of movement, rigidity and gait dysfunction \cite{Jankovic368}. A complication of the mainstay medication used to treat PD, levodopa, is that patients start to experience motor symptom fluctuations related to medication timings. When levodopa is first started, patients experience a smooth and prolonged therapeutic response. As disease progresses (and in a substantial proportion of patients within the first five years), patients start to ''wear off'' from their medications before the next dose, causing a reemergence of parkinsonian symptoms including slowness of gait. These symptom fluctuations impair patients' quality of life and often necessitate changes in medication regime. Motor symptoms can become severe enough to hinder the subject’s gait and movement around their own house \cite{wuehr2020independent}. As a result, the subject may be more likely to stay in one room; once they move, they may typically need more time to transition between rooms. Such outcomes could be used to detect ON and OFF medication motor fluctuations in PD and to inform clinicians and patients of such symptoms. 

Furthermore, a sensitive and accurate ecologically-validated biomarker of PD progression is currently lacking \cite{espay2019}, resulting in multiple failures of clinical trials testing putative neuroprotective therapies in PD \cite{langespay2018,hirschstandaert2021,devos2021}. Gait parameters are sensitive to disease progression symptom change in unmedicated early-stage PD \cite{Zampieri171} and show promise as markers of disease progression \cite{toosizadeh2015motor}, making measuring gait parameters potentially of use in clinical trials of disease-modifying interventions which typically recruit recently diagnosed patients \cite{stephenson2021digital}. Clinical evaluation of PD is normally undertaken in an artificial clinic or laboratory environment where only a snapshot view of the individual’s motor function can be captured. Constant monitoring could capture symptom progression, including motor fluctuations, and sensitively quantify them over time \cite{regnault2019does}. 

Although PD symptoms including gait and balance parameters can be measured continuously at home (with varying degrees of reliability and accuracy) by wearable devices containing inertial motor units (IMUs) or smartphones \cite{Erika2017,di2020technology,s19245465,espay2016}, this data does not show the context in which the measurements are taken (for example where someone is at the time of the symptom). Knowing which room someone is in (indoor localisation) could add valuable holistic information to the interpretation of symptoms of PD. For example, symptoms like freezing of gait \cite{5254658} and turning in gait \cite{giladi2001freezing} vary according to the nature of the setting the person is in, so knowing where someone is could help predict such symptoms or interpret their severity. Furthermore, knowing how much time someone spends alone or with others in a room is a step towards understanding their societal participation \cite{kostanjsek2011use}, which affects quality of life in PD \cite{doi:10.3109/09638288.2010.533245}. Localisation could also add valuable information in the measurement of other behaviors such as non-motor symptoms such as urinary function \cite{https://doi.org/10.1002/mds.27790,hermanowicz2019impact} (e.g. how many times someone visits the toilet room overnight). 

To perform indoor localisation in home environments, IoT-based platforms with sensors capturing various modalities of data combined with machine learning can be used to provide an unobtrusive and continuous localisation \cite{sansano2020multimodal}. Typically, many of these techniques take advantage of the radio-frequency signals, the Received Signal Strength Indication (RSSI), emitted by wearables and measured at access points (AP) throughout a home. These signals to estimate the user’s position from the perceived signal strength, thereby creating radio-map features for each room \cite{8538530}. To provide more accurate localisation, accelerometer data measured by wearable devices, equipped with RSSI measured at receivers, can also be used as it provides a means to distinguish different activities (e.g., walking vs standing). Furthermore, as some activities are tied to particular rooms (e.g. stirring a pan on the hob is very likely to be in a kitchen), accelerometer data may enrich RSSI in differentiating adjacent rooms, which RSSI alone may struggle with \cite{app9204379}.

If accelerometer data are to provide extra features for separating adjacent rooms, greater consideration must be given to data generalisation across different PD patients. As PD is a heterogeneous disease, the symptoms experienced and their severity may vary from one patient to another \cite{https://doi.org/10.1111/ejn.14094}. These severe symptoms, such as tremor, may affect the generalisation of accelerometer data which are prone to bias and accumulated errors \cite{8606925}, especially those worn on the patient's wrists, which is a common and well accepted  placement location \cite{FISHER201644}. Naively combining the accelerometer data with the RSSI may impair the performance of indoor localisation due to differing levels of tremor manifesting in the acceleration signal. In this work, we make two main contributions. 
\begin{enumerate}
    \item We describe the utilisation of RSSI enriched by the accelerometer data to perform room-level localisation. Our proposed network\footnote{Code available at https://github.com/ferdianjovan/Multihead-Dual-Convolutional-Self-Attention} intelligently chooses accelerometer features which may improve the RSSI performance in performing indoor localisation. To properly evaluate our proposed method, we use a free-living (a person living their life freely, without external intervention) dataset created by our group, where the movements and mobility are greatly varied and unstructured as expected in real-world conditions. Our evaluation on such a unique dataset, which includes subjects with and without PD, demonstrates that our proposed network outperforms other approaches in all cross-validation categories.
    \item We also demonstrate how the accurate room-level localisation predictions can be transformed into in-home gait speed biomarkers (e.g. number of room-to-room transition, room-to-room transition duration) which can be used to effectively classify the OFF or ON medication state of a PD patient from this pilot study data.
\end{enumerate} 
\section{Related Work}
\label{sec:related_work}


There has been substantial work using home-based passive sensing systems to assess how the activities and behaviour of people with neurological disease (mainly cognitive dysfunction) change over time \cite{s18082663,info:doi/10.2196/12785}. There is very limited work assessing room use in the home setting in people with Parkinson’s. 

However, gait quantification using wearables or smartphones is an area where a significant amount of work has been done (with several systematic reviews such as these\cite{diseases7010018,es8e622}). Cameras can detect also Parkinsonian gait and some gait features including step length and average walking speed \cite{van2021camera}. Time of flight devices (which measure distances between the subject and the camera \cite{kolb2010}) have been used to assess medication adherence through gait analysis \cite{TUCKER2015120}. From free-living data, one approach to gait and room use evaluation in home settings is by emitting and detecting radio waves to non-invasively track movement. Gait analysis using radio wave technology shows promise to track disease progression, severity and medication response \cite{9772932}. However, this approach cannot identify who is doing the movement and also suffers from technical issues when the radio waves are occluded by another object. Much of the work done so far using video to track PD symptoms has focused on the performance of structured clinical rating scales during telemedicine consultations as opposed to naturalistic behaviour \cite{sibley2021video}, and there have been some privacy concerns around the use of video data at home \cite{devries2019}.

RSSI data produced from wearable devices is a type of data with fewer privacy concerns; it can be measured continuously and unobtrusively over long periods of time to capture real-world function and behavior in a privacy-friendly way. In indoor localisation, fingerprinting using RSSI is the typical technique used to estimate the wearable (user) location by using signal strength data representing a coarse and noisy estimate of the distance access point from the wearable \cite{s17010036,Schindhelm2011}. RSSI signals are not stable, they fluctuate randomly due to shadowing, fading and multi-path effects. However, many techniques have been proposed in recent years to tackle these fluctuations, and, indirectly, improve the localisation accuracy.  Some of the works \cite{ZHANG2016279} utilise deep neural networks (DNN) to generate coarse positioning estimates from RSSI signals, which are then refined by a hidden Markov model (HMM) to produce a final estimate location. Other works, \cite{8538530}, try to utilise a time-series of RSSI data and exploit the temporal connections within each access point to estimate room-level position. A CNN is used to build localisation models to further leverage the temporal dependencies across time-series readings.

It has been suggested that we cannot rely on RSSI alone for indoor localisation in home environments for PD subjects due to shadowing rooms with tight separation \cite{pandey2021,app9204379,sansano2020multimodal}. Sansano et al. combine RSSI signals and inertial measurement unit (IMU) data to test the viability of leveraging other sensors in aiding the positioning system to produce a more accurate location estimate \cite{sansano2020multimodal}. Classic machine learning approaches such as Random Forest (RF), Artificial Neural Network (ANN), k-Nearest Neighbour (k-NN) are tested, and the result shows that the RF outperforms other methods in tracking a person in indoor environments. Poulose et al. combine smartphone IMU sensor data and Wi-Fi received signal strength indication (RSSI) measurements to estimate the exact location (in Euclidean position X, Y) of a person in indoor environments \cite{app9204379}. The proposed sensor fusion framework uses location fingerprinting in combination with a pedestrian dead reckoning (PDR) algorithm to reduce the positioning errors. 

Looking at this multi-modality classification / regression problem from a timeseries perspective, there has been a lot of explorations in tackling a problem where each modality can be categorised as multivariate timeseries data \cite{DTML2021,NEURIPS2020_cdf6581c,TFT2019}. LSTM and attention layers are often used in parallel to directly transform raw multivariate time series data into low-dimensional feature representation for each modality. Later, various processed is done to further extract correlations across modalities through the use of various layers (e.g. concatenation, CNN layer, transformer, self-attention) \cite{DTML2021, TFT2019}. Our work is inspired by Sansano-Sansano et al. \cite{sansano2020multimodal} where we only utilise accelerometer data to enrich the RSSI, instead of utilising all IMU sensors, in order to reduce battery consumption. In addition, unlike Sansano-Sansano et al. who stop at predicting room locations, we go a step further and use room-to-room transition behaviours, as features for a binary classifier predicting whether people with PD are taking their medications or withholding them. 



\section{Cohort and Dataset}
\label{sec:cohort}

\begin{figure}[t!]
	\centering
	\includegraphics[width=0.46\textwidth]{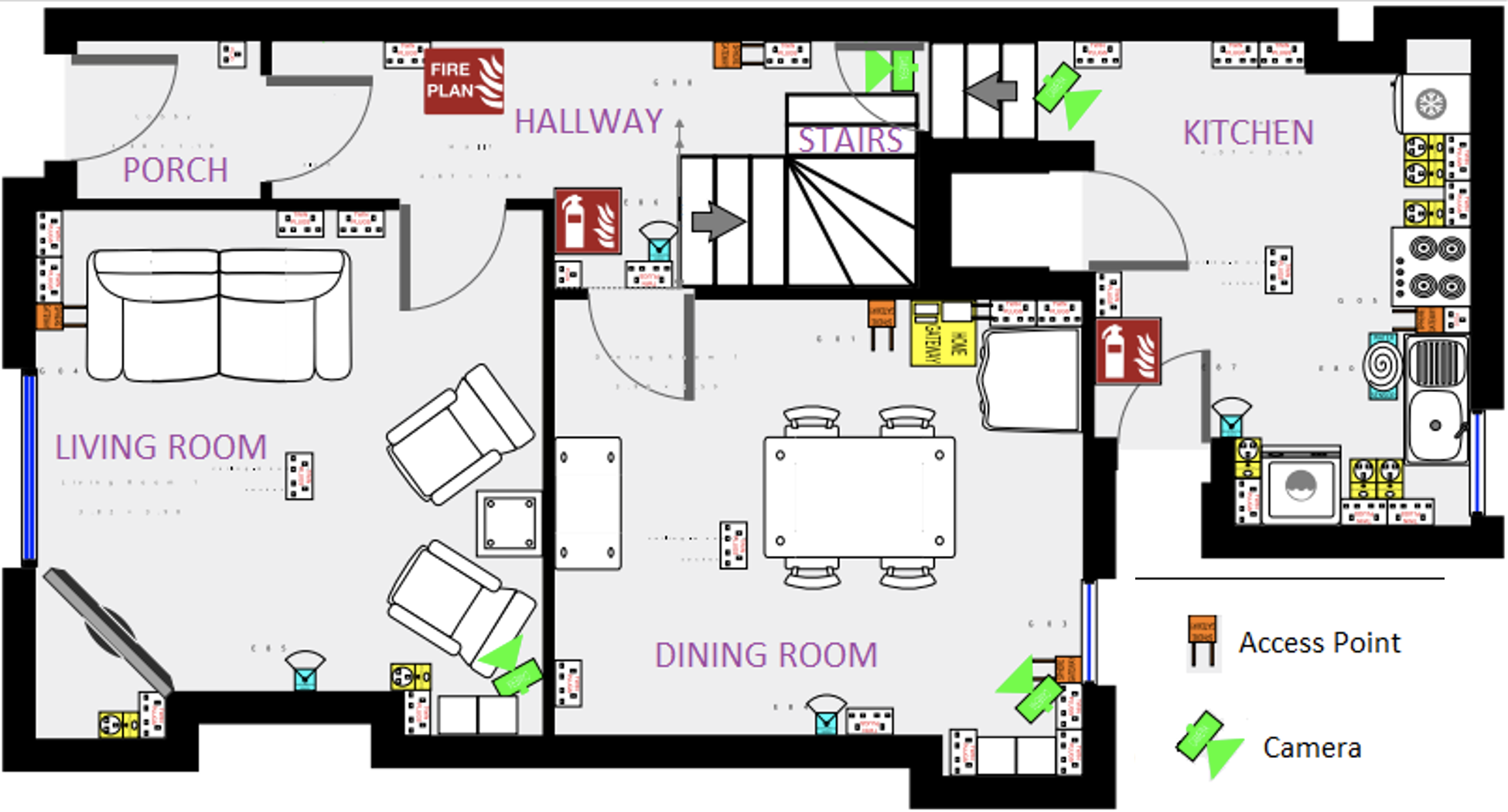}
	\caption{Layout of the residential home setting.} 
	\label{fig:home_layout}
\end{figure}

\textbf{Dataset.}  This dataset was collected using wristband wearable sensors, one on each wrist of all participants, containing tri-axial accelerometers\footnote{The wearables are custom-designed and purposefully do not utilise gyroscope and magnetometer sensors to increase battery life.} and 10 Access Points (APs) placed through the residential home (see Fig. \ref{fig:home_layout} for house layout and AP location), each measuring the RSSI \cite{8369009}. The wearable devices wirelessly transmit data using the Bluetooth Low Energy (BLE) standard which can be received by the 10 APs. Each AP records the transmitted packets from the wearable sensor which contains the accelerometer readings sampled at 30Hz, with each AP recording RSSI values sampled at 5 Hz. 

The dataset contains 12 spousal/parent-child/friend-friend pairs (24 participants in total) living freely in a smart home for five days. Each pair consists of one person with PD and one person as the healthy control volunteer (HC). This pairing was chosen to enable PD vs HC comparison, for safety reasons and also to increase the naturalistic social behaviour (particularly amongst the spousal pairs who already lived together). From the 24 participants, five females and seven males have PD. The average age of the participants is 60.25 (PD 61.25, Control 59.25) and the average time since PD diagnosis for the person with PD is 11.3 years (range 0.5-19). 

To measure the accuracy of the machine learning models, wall-mounted cameras are installed in the ground floor of the house which capture red-green-blue (RGB) and depth data 2-3 hours daily (during daylight hours at times when participants were at home). The videos were then manually annotated to the nearest millisecond to provide localisation labels. Multiple human labellers used a widely available software called ELAN \cite{aguera2011elan} to watch up to 4 simultaneously-captured video files at a time. The resulting labelled data recorded the kitchen, hallway, dining room, living room, stairs, and porch. The duration of labelled data recorded by the cameras for PD and HC is 72.84 and 75.31 hours, respectively, which provides a relatively balanced label set for our room-level classification\footnote{Approaches generalise to all rooms over the entire period, however we limit the time frame and rooms to only those that were manually annotated (on ground floor).}. Finally, to evaluate the ON/OFF medication state, participants with PD were asked to withhold their dopaminergic medications so that they were in the practically-defined OFF medications state for a temporary period of several hours during the study. Withholding medications removes their mitigation on symptoms, leading to mobility deterioration which can include slowing of gait. 

\textbf{Data pre-processing for indoor localisation.} The data from the two wearable sensors worn by each participant were combined at each time point, based on their modality, i.e. twenty RSSI values (corresponding to 10 APs for each of the two wearable sensors), and accelerometry traces in six spatial directions (corresponding to the three spatial directions (x, y, z) for each wearable) were recorded at each time point. The accelerometer data is resampled to 5Hz to synchronise the data with RSSI values. With a 5-second time window and 5Hz sampling rate, each RSSI data sample has an input of size (25 x 20) and accelerometer data has an input of size (25 x 6). Imputation for missing values, specifically for RSSI data, is applied by replacing the missing values with a value that is not possible normally (i.e., -120dB). Missing values exist in RSSI data whenever the wearable is out of range of an AP. Finally, all time-series measurements by the modalities are normalised. 

\textbf{Data pre-processing for medication state.} Our main focus is for our neural network to continuously produce room predictions which are then transformed into in-home gait speed features, particularly for persons with PD. We hypothesise that during their OFF medication state, the deterioration in mobility of a person with PD is exhibited by how they transition between rooms. These features include `Room-to-room Transition Duration', and the `Number of Transitions' between two rooms. `Number of Transitions' represents how active PD subjects are within a certain period of time, while `Room-to-room Transition Duration' may provide insight into how severe their disease is by the speed with which they navigate their home environment. With the layout of the house where participants stayed (see Fig. \ref{fig:home_layout}), the hallway is used as a hub connecting all other rooms labelled, and `Room-to-room Transition' shows the transition duration (in seconds) between two rooms connected by the hallway. The transition between (1) kitchen and living room, (2) kitchen and dining room, and (3) dining room and living room are chosen as the features due to their commonality across all participants. For these features, we limit the transition time duration (i.e. the time spent in the hallway) to 60 seconds to exclude transitions likely to be prolonged and thus may not be representative of the person's mobility.

These in-home gait speed features are produced by an indoor-localisation model by feeding RSSI signals and accelerometer data from 12 PD participants from 6 a.m. to 10 p.m. daily which are aggregated into 4 hour windows. From this, each PD participant will have 20 data samples (four data samples for each of the five days), each of which contains six features (three for the mean of room-to-room transition duration, and three for the number of room-to-room transitions). There is only one 4-hour window during which the person with PD is OFF medications. These samples are then used to train a binary classifier\footnote{The Random Forest is chosen as a binary classifier; It is not the one used for indoor localisation.} determining whether a person with PD is ON or OFF their medications.  

For a baseline comparison to the in-home gait speed features, demographic features which include age, gender, years of PD, and MDS-UPDRS III score (the gold-standard clinical rating scale score used in clinical trials to measure motor disease severity in PD) are chosen. Two MDS-UPDRS III scores are assigned for each PD participant; one is assigned when a person with PD is ON medications, and the other one is assigned when a person with PD is OFF medications. For each in-home gait speed feature data sample, there will be a corresponding demographic feature data sample which are used to train a different binary classifier to predict whether a person with PD is ON or OFF medications. 


\textbf{Ethical approval.}  Full approval from NHS Wales Research Ethics Committee 6 was granted on 17$^{th}$ December 2019, and Health Research Authority and Health and Care Research Wales approval confirmed on 14$^{th}$ January 2020; the research was conducted in accord with the Helsinki Declaration of 1975; written informed consent was gained from all study participants. In order to protect participant privacy supporting data is not shared openly. It will be made available to bona fide researchers subject to a data access agreement. If you wish to apply to access this data, please email data-bris@bristol.ac.uk.

\section{Methodologies and Framework}
\label{sec:proposed_framework}

\begin{figure}[t!]
	\centering
	\includegraphics[width=0.46\textwidth]{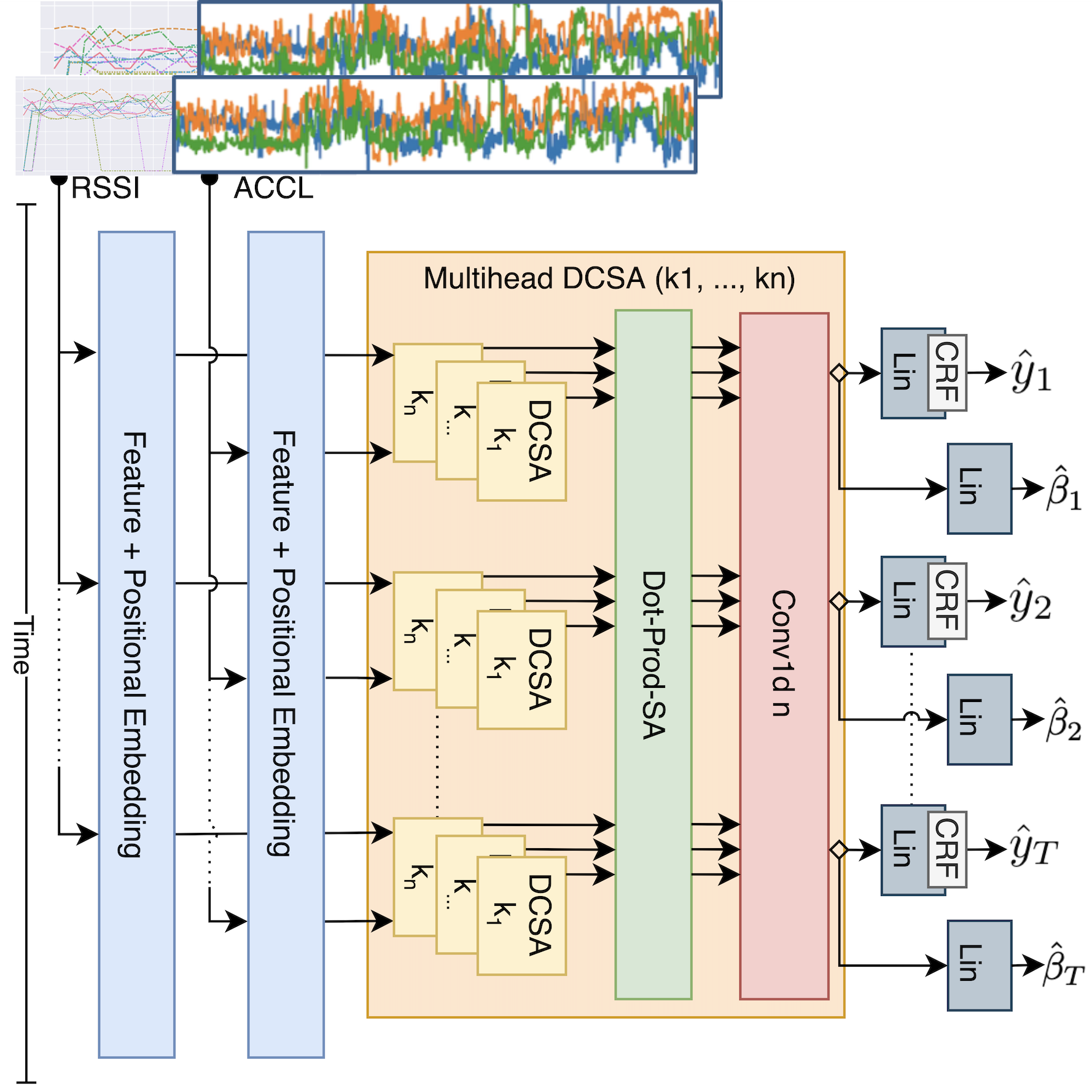}
        \caption{MDCSA architecture.}
	\label{fig:mdcsa_architecture}
\end{figure}

We introduce Multihead Dual Convolutional Self Attention (MDCSA), a deep neural network that utilises dual modalities for indoor localisation in home environments. The network tackles two challenges that arise from multimodality and time-series data: 
\begin{enumerate}
    \item \textbf{Capturing multivariate features and filtering multimodal noises.}  RSSI signals, which are measured at multiple access points within a home received from wearable communication, have been widely used for indoor localisation~\cite{8538530}, typically using a fingerprinting technique that produces a ground truth radio map of a home. Naturally, the wearable also produces acceleration measurements which can be used to identify typical activities performed in a specific room, and thus we can explore if accelerometer data will enrich the RSSI signals, in particular to help distinguish adjacent rooms, which RSSI-only systems typically struggle with. If it will, how can we incorporate these extra features (and modalities) into the existing features for accurate room predictions, particularly in the context of PD where the acceleration signal may be significantly impacted by the disease itself?
    \item \textbf{Modelling local and global temporal dynamics.} The true correlations between inputs both intra-modality (i.e. RSSI signal among access points) and inter-modality (i.e. RSSI signal against accelerometer fluctuation) are dynamic. These dynamics can affect one another within a local context (e.g. cyclical patterns) or across long-term relationships. Can we capture local and global relationships across different modalities?
\end{enumerate}

The MDCSA architecture, shown in Figure \ref{fig:mdcsa_architecture}, addresses the aforementioned challenges through a series of neural network layers which are described in the following sections.

\subsection{Modality Positional Embedding}
\label{subsec:pos_embed}

Due to different data dimensionality between RSSI and accelerometer, coupled with the missing temporal information, a linear layer with a positional encoding is added to transform both RSSI and accelerometer data into their respective embeddings. Suppose we have a collection of RSSI signals $\mathbf{x}^r = [\mathbf{x}_1^r, \ldots, \mathbf{x}_T^r] \in \mathbb R^{T \times r}$, and accelerometer data $\mathbf{x}^a = [\mathbf{x}_1^a, \ldots, \mathbf{x}_T^a] \in \mathbb R^{T \times a}$ within $T$ time unit, where $\mathbf{x}^r_t = \big[x^1_t, \ldots, x^r_t\big]$ represents RSSI signals from $r$ access points, and $\mathbf{x}^a_t = \big[x^1_t, \ldots, x^a_t\big]$ represents accelerometer data from $a$ spatial directions at time $t$ with $t \leq T$. Given feature vectors $\mathbf{x}^u_t = \big[x^1_t, \ldots, x^u_t\big]$ with $u \in \{r, a\}$ representing RSSI or accelerometer data at time $t$, and $t \leq T$ representing time index, a positional embedding $\mathbf{h}^u_t$ for RSSI or accelerometer can be obtained by:
\begin{equation}
	\label{eq:linear_embedding}
	\mathbf{h}^u_t = \big(\mathbf W_u  \mathbf{x}^u_t + \mathbf b_u\big) + \overrightarrow{p_t}
\end{equation}
\noindent where $\mathbf W_u \in \mathbb R^{u \times d}$, and $\mathbf b_u \in \mathbb R^{d}$ are weight and bias to learn, $d$ is the embedding dimension, and $\overrightarrow{p_t} \in \mathbb R^{d}$ is the corresponding position encoding at time $t$.

\subsection{Locality Enhancement with Self-Attention}
\label{subsec:local_enhancement}


As it is time series data, the importance of an RSSI or accelerometer value at each point in time can be identified in relation to its surrounding values - such as cyclical patterns, trends, or fluctuations. Utilising historical context that can capture local patterns on top of point-wise values, performance improvements in attention-based architectures can be achieved. One straightforward option is to utilise a recurrent neural network such as a long-short term memory (LSTM) approach. However, in LSTM layers, the local context is summarised based on the previous context and the current input. Two similar patterns separated by a long period of time might have different context if they are processed by the LSTM layers \cite{https://doi.org/10.48550/arxiv.1409.0473}. We utilise a combination of \textit{causal convolution} layers and self-attention layers which we name Dual Convolutional Self-Attention (DCSA). The DCSA takes in a primary input $\mathbf{\hat{x}_1} \in \mathbb{R}^{T \times d}$ and a secondary input $\mathbf{\hat{x}_2} \in \mathbb{R}^{T \times d}$ and yields:
\begin{equation}
	\label{eq:dcsa}
	\begin{aligned}
	DCSA_k\big(\mathbf{\hat{x}_1}, \mathbf{\hat{x}_2}\big) = GRN\big( & Norm(\Phi(\mathbf{\hat{x}_1}) + \mathbf{\hat{x}_1}), \\ & Norm(\Phi(\mathbf{\hat{x}_2}) + \mathbf{\hat{x}_2})\big)
	\end{aligned}
\end{equation}
\noindent with
\begin{equation}
	\label{eq:csa}
		\Phi\big(\mathbf{\hat{x}}\big) = SA\Big(\Phi_k(\mathbf{\hat{x}})~\mathbf W_{\mathbf Q}, \Phi_k(\mathbf{\hat{x}})~\mathbf W_{\mathbf K}, \mathbf{\hat{x}}~\mathbf W_{\mathbf V}\Big)
\end{equation}
\noindent where $GRN(.)$ is Gated Residual Network (GRN), introduced in \cite{TFT2019}, to integrate dual inputs into one integrated embedding, $Norm(.)$ is a standard layer normalisation, $SA(.)$ is a scaled dot-product self-attention introduced in \cite{NIPS2017_3f5ee243}, $\Phi_k(.)$ is a 1D-convolutional layer with a kernel size $\{1, k\}$ and a stride 1, $\mathbf W_{\mathbf K} \in \mathbb R^{d \times d}$, $\mathbf W_{\mathbf Q} \in \mathbb R^{d \times d}$ , $W_{\mathbf V} \in \mathbb R^{d \times d}$ are weights for keys, queries and values of the self-attention layer, and $d$ is the embedding dimension. Note that all weights for GRN are shared across each time step $t$.

\subsection{Multihead Dual Convolutional Self-Attention}

Our approach employs a self-attention mechanism introduced in \cite{NIPS2017_3f5ee243} to capture global dependencies across time steps. It is embedded as part of the DCSA architecture. Inspired by Vaswani et al. \cite{NIPS2017_3f5ee243} in utilising multihead self-attention, we utilise our DCSA with various kernel lengths with the same aim: allowing asymmetric long-term learning. The multihead DCSA, shown as part in Figure \ref{fig:mdcsa_architecture}, takes in two inputs $\mathbf{\hat{x}_1}, \mathbf{\hat{x}_2} \in \mathbb{R}^{T \times d}$ and yields:
\begin{equation}
	\label{eq:mdcsa}
	\begin{aligned}
		MDCSA_{k_1, \ldots, k_n}\big(\mathbf{\hat{x}_1}, \mathbf{\hat{x}_2}\big) = \Phi_n\big(\Phi_{k_1, \ldots, k_n}(\mathbf{\hat{x}_1}, \mathbf{\hat{x}_2})\big)
	\end{aligned}
\end{equation}
\noindent with
\begin{equation}
	\label{eq:outer_sa}
	\begin{aligned}
		\Phi_{k_1, \ldots, k_n}\big(\mathbf{\hat{x}_1}, \mathbf{\hat{x}_2}\big) = SA\big( & \Xi_{k_1, \ldots, k_n}(\mathbf{\hat{x}_1}, \mathbf{\hat{x}_2})~\mathbf W_{\mathbf Q}, \\
		& \Xi_{k_1, \ldots, k_n}(\mathbf{\hat{x}_1}, \mathbf{\hat{x}_2})~\mathbf W_{\mathbf K}, \\ 
		& \Xi_{k_1, \ldots, k_n}(\mathbf{\hat{x}_1}, \mathbf{\hat{x}_2})~\mathbf W_{\mathbf V}\big)
	\end{aligned}
\end{equation}
\begin{equation}
	\label{eq:multidcsa}
	\begin{aligned}
		\Xi_{k_1, \ldots, k_n}\big(\mathbf{\hat{x}_1}, \mathbf{\hat{x}_2}\big) = \big[ & DCSA_{k_1}(\mathbf{\hat{x}_1}, \mathbf{\hat{x}_2}), \ldots, DCSA_{k_n}(\mathbf{\hat{x}_1}, \mathbf{\hat{x}_2})\big]
	\end{aligned}
\end{equation}
\noindent where $\Phi_n(.)$ is a 1D-convolutional layer with a kernel size $\{1, n\}$ and a stride $n$, $\mathbf W_{\mathbf K} \in \mathbb R^{d \times d}$, $\mathbf W_{\mathbf Q} \in \mathbb R^{d \times d}$ , $W_{\mathbf V} \in \mathbb R^{d \times d}$ are weights for keys, queries and values of the self-attention layer, and $\Xi_{k_1, \ldots, k_n}(.)$ concatenates the output of each $DCSA_k(.)$ in temporal order. For regularisation, a normalisation layer followed by a dropout layer is added after Equation \ref{eq:mdcsa}.

Following the modality positional embedding layer in subsection \ref{subsec:pos_embed}, the positional embeddings of RSSI $\mathbf{h}^r = [\mathbf{h}^r_1, \ldots, \mathbf{h}^r_T]$ and accelerometer $\mathbf{h}^a = [\mathbf{h}^a_1, \ldots, \mathbf{h}^a_T]$, produced by Eq. \ref{eq:linear_embedding}, are then fed to an MDCSA layer with various kernel sizes $[k_1, \ldots, k_n]$:
\begin{equation}
    \label{eq:pe_to_mdcsa}
    \begin{aligned}
        \boldsymbol{\mathfrak h} = MDCSA_{k_1, \ldots, k_n}\big(\mathbf{h}^r, \mathbf{h}^a\big)
    \end{aligned}
\end{equation}
\noindent to yield $\boldsymbol{\mathfrak h} = \big[\mathfrak h_1, \ldots, \mathfrak h_T\big]$ with $\mathfrak h_t \in \mathbb R^{d}$ and $t \leq T$.

\subsection{Final Layer and Loss Calculation}
\label{subsec:nonlinear}

We apply two different layers to produce two different outputs during training. The room-level predictions are produced via a single conditional random field (CRF) layer in combination with a linear layer applied to the output of Eq. \ref{eq:pe_to_mdcsa} to produce the final predictions as
\begin{equation}
	\hat y_t = CRF\big( \Phi( \mathfrak h_t ) \big)
\end{equation}
\begin{equation}
    \label{eq:hidden2loc}
    \Phi\big( \mathfrak h_t \big) = \mathbf W_p \mathfrak h_t + \mathbf b_p
\end{equation}
\noindent where $\mathbf W_p \in \mathbb R^{d \times m}$, and $\mathbf b_p \in \mathbb R^{m}$ are weight and bias to learn, $m$ is the number of room locations, and $\boldsymbol{\mathfrak h} = \big[\mathfrak h_1, \ldots, \mathfrak h_T\big] \in \mathbb R^{T \times d}$ is the refined embedding produced by Eq. \ref{eq:pe_to_mdcsa}. Even though the transformer can take into account neighbour information before generating the refined embedding at time step $t$, its decision is independent; it does not take into account the actual decision made by other refined embeddings $t$. We use a CRF layer to cover just that, i.e. to maximise the probability of the refined embeddings of all time steps, so it can better model cases where refined embeddings closest to one another must be compatible (i.e. minimising the possibility for impossible room transitions). When finding the best sequence of room location $\hat y_t$, the Viterbi Algorithm is used as a standard for the CRF layer.

For the second layer, we choose a particular room as a reference and perform a binary classification at each time step $t$. The binary classification is produced via a linear layer applied to the refined embedding $\boldsymbol{\mathfrak h}$ as
\begin{equation}
	 \hat{\beta}_t = \mathbf W_{\beta}  \mathfrak h_t + \mathbf b_{\beta}
\end{equation}
\noindent where $\mathbf W_{\beta} \in \mathbb R^{d \times 1}$, and $\mathbf b_{\beta} \in \mathbb R$ are weight and bias to learn, and $\hat{\beta} = \big[\hat{\beta}_1, \ldots, \hat{\beta}_T\big] \in \mathbb R^T$ is the target probabilities for the referenced room within time window $T$. The reason to perform a binary classification against a particular room is because of our interest in improving the accuracy in predicting that room. In our application, the room of our choice is the hallway where it will be used as a hub connecting any other room. 

\textbf{Loss Functions.} During the training process, the MDCSA network produces two kinds of outputs. Emission outputs (outputs produced by Equation \ref{eq:hidden2loc} prior to prediction outputs) $\mathbf{\hat{e}} = \Big[\Phi\big( \mathfrak h_1 \big), \ldots, \Phi\big( \mathfrak h_T \big)\Big]$ are trained to generate the likelihood estimate of room predictions, while the binary classification output $\hat{\beta} = [\hat{\beta}_1, \ldots, \hat{\beta}_T]$ is used to train the probability estimate of a particular room. The final loss function can be formulated as a combination of both likelihood and  binary cross entropy loss function described as:
\begin{equation}
	\mathcal L(\mathbf{\hat{e}}, \mathbf y, \hat \beta, \beta) = \mathcal L_{NLL}(\mathbf{\hat{e}}, \mathbf y) + \displaystyle\sum_{i=1}^T~\mathcal L_{BCE}(\hat \beta_i, \beta_i)
\end{equation}
\begin{equation}
    \begin{aligned}
        \mathcal L_{NLL}(\mathbf{\hat{e}}, \mathbf y) = & \displaystyle\sum_{\hat y} \displaystyle\sum_{i=0}^{T} P\big(\Phi\big( \mathfrak h_i \big) \mid \hat y_i\big) T\big(\hat y_i \mid \hat y_{i-1}\big) - \\ 
        & \displaystyle\sum_{i=0}^{T} P\big(\Phi\big( \mathfrak h_i \big) \mid y_i\big)T\big(y_i \mid  y_{i-1}\big)
    \end{aligned}
\end{equation}
\begin{equation}
	\label{eq:bce}
	\mathcal L_{BCE}(\hat \beta, \beta) = -\frac{1}{T} \displaystyle\sum_{i=0}^{T} \beta_i~log(\hat{\beta}_i) + (1-\beta_i)~log(1 - \hat{\beta}_i)
\end{equation}
\noindent where $\mathcal L_{NLL}(.)$ represents the negative log-likelihood and $\mathcal L_{BCE}(.)$ denotes the binary cross entropy, $\mathbf y = [y_1, \ldots, y_T] \in \mathbb R^T$ is the actual room locations, and $\beta = [\beta_1, \ldots, \beta_T] \in \mathbb R^T$ is the binary value whether at time $t$ the room is the referenced room or not. $P(x \mid y)$ denotes the conditional probability, and $T(y_i \mid y_{i-1})$ denotes the transition matrix cost of having transitioned from $y_{i-1}$ to $y$.
\section{Experiments and Results}
\label{sec:experiment}

\begin{table*}[t!]
	\centering
        \caption{Room-level and medication state accuracy of all models. Standard deviation is shown in $(.)$, the best performer is bold, while the second best is italicized. Note that our proposed model is the one named MDCSA$_{1,4,7}$}
	\begin{tabular}{cccccc}
		\toprule[1pt]
		\multirow{2}{*}{\textbf{Training}} & \multirow{2}{*}{\textbf{Model}} & \multicolumn{2}{c}{\textbf{Room-Level Localisation}} & \multicolumn{2}{c}{\textbf{Medication State}}\\
		& & \textbf{Precision} & \textbf{F1-Score} & \textbf{F1-Score} & \textbf{AUROC} \\ 
		\midrule[0.5pt]
		\multirow{7}{*}{\textbf{ALL-HC}} & RF & \textbf{95.00} & \textbf{95.20} & 56.67 (17.32) & \textbf{84.55 (12.06)} \\
		& TENER & 94.60 &  94.80 & 47.08 (16.35) & 67.74 (10.82) \\
		& DTML & 94.80 & 94.90 & 50.33 (13.06) & 75.97 (9.12) \\
		& Alt DTML & 94.80 & 95.00 & 47.25 (5.50) & 75.63 (4.49) \\
		& MDCSA$_{1,4,7}$ 4APS & 92.22 & 92.22 & 53.47 (12.63) & 73.48 (6.18) \\
		& MDCSA$_{1,4,7}$ RSSI & 94.70 & 94.90 & 51.14 (11.95) & 68.33 (18.49) \\
		& MDCSA$_{1,4,7}$ 4APS RSSI & 93.30 & 93.10 & \textbf{64.52 (11.44)} & \textit{81.84 (6.30)} \\
		& MDCSA$_{1,4,7}$ & \textit{94.90} & \textit{95.10} & \textit{64.13 (6.05)} & 80.95 (10.71) \\ [0.6ex]
		\cdashline{2-6} \\ [-2ex]
		& \multicolumn{3}{l}{\textbf{Demographic Features}} & 49.74 (15.60) & 65.66 (18.54) \\
		\midrule[0.5pt]
		\multirow{7}{*}{\textbf{LOO-HC}} & RF & 89.67 (1.85) & 88.95 (2.61) & 54.74 (11.46) & 69.24 (17.77) \\
		& TENER & 90.35 (1.87) & 89.75 (2.24) & 51.76 (14.37) & 70.80 (9.78) \\
		& DTML & 90.51 (1.95) & \textit{89.82 (2.60)} & 55.34 (13.67) & 73.77 (9.84) \\
		& Alt DTML & \textit{90.52 (2.17)} & 89.71 (2.83) & 49.56 (17.26) & 73.26 (10.65) \\
		& MDCSA$_{1,4,7}$ 4APS & 88.01 (6.92) & 88.08 (5.73) & \textbf{59.52 (20.62)} & 74.35 (16.78) \\
		& MDCSA$_{1,4,7}$ RSSI & 90.26 (2.43) & 89.48 (3.47) & \textit{58.84 (23.08)} & \textit{76.10 (10.84)} \\
		& MDCSA$_{1,4,7}$ 4APS RSSI & 88.55 (6.67) & 88.75 (5.50) & 42.34 (13.11) & 72.58 (6.77) \\
		& MDCSA$_{1,4,7}$ & \textbf{91.39 (2.13)} & \textbf{91.06 (2.62)} & 55.50 (15.78) & \textbf{83.98 (13.45)} \\ [0.6ex]
		\cdashline{2-6} \\ [-2ex]
		& \multicolumn{3}{l}{\textbf{Demographic Features}} & 51.79 (15.40) & 68.33 (18.43) \\		
		\midrule[0.5pt]
		\multirow{7}{*}{\textbf{LOO-PD}} & RF & 86.89 (7.14) & 84.71 (7.33) & 43.28 (14.02) & 62.63 (20.63) \\
		& TENER & 86.91 (6.76) & 86.18 (6.01) & 36.04 (9.99) & 60.03 (10.52) \\
		& DTML & 87.13 (6.53) & 86.31 (6.32) & 43.98 (14.06) & 66.93 (11.07) \\
		& Alt DTML & 87.36 (6.30) & 86.44 (6.63) & 44.02 (16.89) & 69.70 (12.04) \\
		& MDCSA$_{1,4,7}$ 4APS & 86.44 (6.96) & 85.93 (6.05) & \textit{47.26 (14.47)} & \textit{72.62 (11.16)} \\
		& MDCSA$_{1,4,7}$ RSSI & \textit{87.61 (6.64)} & \textit{87.21 (5.44)} & 45.71 (17.85) & 67.76 (10.73) \\
		& MDCSA$_{1,4,7}$ 4APS RSSI & 87.20 (7.17) & 87.00 (6.12) & 41.33 (17.72) & 66.26 (12.11) \\
		& MDCSA$_{1,4,7}$ & \textbf{88.04 (6.94)} & \textbf{87.82 (6.01)} & \textbf{49.99 (13.18)} & \textbf{81.08 (8.46)} \\ [0.6ex]
		\cdashline{2-6} \\ [-2ex]
		& \multicolumn{3}{l}{\textbf{Demographic Features}} & 43.89 (14.43) & 60.95 (25.16) \\		
		\midrule[0.5pt]
		\multirow{7}{*}{\textbf{4m-HC}} & RF & 74.27 (8.99) & 69.87 (7.21) & \textit{50.47 (12.63)} & 59.55 (12.38) \\
		& TENER & 69.86 (18.68) & 60.71 (24.94) & N/A & N/A \\
		& DTML & 77.10 (9.89) & 70.12 (14.26) & 43.89 (11.60) & 64.67 (12.88) \\
		& Alt DTML & 78.79 (3.95) & 71.44 (9.82) & 47.49 (14.64) & 65.16 (12.56) \\
		& MDCSA$_{1,4,7}$ 4APS & 81.42 (6.95) & 78.65 (7.59) & 42.87 (17.34) & 67.09 (7.42) \\
		& MDCSA$_{1,4,7}$ RSSI & 81.69 (6.85) & 77.12 (8.46) & 49.95 (17.35) & \textit{69.71 (11.55)} \\
		& MDCSA$_{1,4,7}$ 4APS RSSI & \textit{82.80 (7.82)} & \textit{79.37 (8.98)} & 43.57 (23.87) & 65.46 (15.78) \\
		& MDCSA$_{1,4,7}$ & \textbf{83.32 (6.65)} & \textbf{80.24 (6.85)} & \textbf{55.43 (10.48)} & \textbf{78.24 (6.67)} \\ [0.6ex]
		\cdashline{2-6} \\ [-2ex]
		& \multicolumn{3}{l}{\textbf{Demographic Features}} & 32.87 (13.81) & 53.68 (13.86) \\		
		\midrule[0.5pt]
		\multirow{7}{*}{\textbf{4m-PD}} & RF & 71.00 (9.67) & 65.89 (11.96) & N/A & N/A \\
		& TENER & 65.30 (23.25) & 58.57 (27.19) & N/A & N/A \\
		& DTML & 70.35 (14.17) & 64.00 (17.88) & N/A & N/A \\
		& Alt DTML & 74.43 (9.59) & 67.55 (14.50) & N/A & N/A \\
		& MDCSA$_{1,4,7}$ 4APS & 81.02 (8.48) & \textit{76.85 (10.94)} & \textbf{49.97 (7.80)} & \textit{69.10 (7.64)} \\
		& MDCSA$_{1,4,7}$ RSSI & 77.47 (12.54) & 73.99 (13.00) & 41.79 (16.82) & 67.37 (16.86) \\
		& MDCSA$_{1,4,7}$ 4APS RSSI & \textit{83.01 (6.42)} & \textbf{79.77 (7.05)} & 41.18 (12.43) & 63.16 (11.06) \\
		& MDCSA$_{1,4,7}$ & \textbf{83.30 (6.73)} & 76.77 (13.19) & \textit{48.61 (12.03)} & \textbf{76.39 (12.23)} \\ [0.6ex]
		\cdashline{2-6} \\ [-2ex]
		& \multicolumn{3}{l}{\textbf{Demographic Features}} & 36.69 (18.15) & 50.53 (15.60) \\			
		\bottomrule[1pt]
	\end{tabular}
        \label{tab:benchmark_results}
\end{table*}

We compare our proposed network, MDCSA$_{1, 4, 7}$\footnote{We drop the `$_{1, 4, 7}$' part when the context is clear.} (MDCSA with 3 kernels of size $1$, $4$, and $7$), with:
\begin{itemize}
    \item Random Forest (RF) as a baseline technique which has been shown to work well for indoor localisation \cite{thanaphon2022},
    \item TENER \cite{yan2019tener} which is a modified transformer encoder in combination with a CRF layer representing a model with capability to capture global dependency and enforce dependencies in temporal aspects,
    \item DTML \cite{DTML2021} represents the state-of-the-art model for multimodal and multivariate time series with a transformer encoder to learn asymmetric correlations across modalities,
    \item Alt DTML\footnote{Our attempt to see the effect of GRN and CRF layer on a SOTA model.} representing DTML with a GRN layer replacing the context aggregation layer and CRF layer added as the last layer, 
    \item MDCSA$_{1, 4, 7}$ 4APS, as an ablation study, with our proposed network (i.e. MDCSA$_{1, 4, 7}$) using 4 access points for the RSSI (instead of 10 access points) and accelerometer data (ACCL) as its input features,
    \item MDCSA$_{1, 4, 7}$ RSSI, as an ablation study, with our proposed network using only RSSI, without ACCL, as its input features, and
    \item MDCSA$_{1, 4, 7}$ 4APS RSSI, as an ablation study, with our proposed network using only 4 access points for the RSSI as its input features.
\end{itemize}
For RF, all the time series features of RSSI and accelerometry are flattened and merged into one feature vector for room-level localisation. For TENER, at each time step $t$, RSSI $\mathbf x^r_t$ and accelerometer $\mathbf x^a_t$ features are combined via a linear layer before they are processed by the networks. A grid search on the parameters of each network is performed to find the best parameter for each model. The parameters to tune are: the embedding dimension $d$ in $\{128, 256\}$, the number of epochs in $\{200, 300\}$, and the learning rate in $\{0.01, 0.0001\}$. The dropout rate is set to 0.15, and the RAdam optimiser \cite{radam2019} in combination with Look-Ahead algorithm \cite{lookahead2019} is used for the training with early stopping using the validation performance. For the RF, we perform a cross-validated parameter search for the number of trees ($\{200, 250\}$), the minimum number of samples in a leaf node ($\{1, 5\}$), and whether a warm start is needed ($\{True, False\}$). The Gini impurity is used to measure splits. 

\begin{table}[t!]
	\centering
	\setlength\tabcolsep{2.5pt}
        \caption{Hallway prediction on limited training data.}
        \begin{tabular}{cccc}
		\toprule[1pt]
		\textbf{Training} & \textbf{Model} & \textbf{Precision} & \textbf{F1-Score} \\ 
		\midrule[0.5pt]
		\multirow{3}{*}{\textbf{4m-HC}} & MDCSA 4APS RSSI & 62.32 (19.72) & 58.99 (23.87) \\
		& MDCSA 4APS & 68.07 (23.22) & 60.01 (26.24) \\
		& MDCSA & 71.25 (21.92) & 68.95 (17.89) \\
		\midrule
		\multirow{3}{*}{\textbf{4m-PD}} & MDCSA 4APS RSSI & 58.59 (23.60) & 57.68 (24.27) \\
		& MDCSA 4APS & 62.36 (18.98) & 57.76 (20.07) \\
		& MDCSA & 70.47 (14.10) & 64.64 (21.38) \\
		\bottomrule[1pt]
	\end{tabular}
	\label{tab:hallway_result}
\end{table}

\textbf{Evaluation Metrics.} We are interested in developing a system to monitor PD motor symptoms in home environments. For example, we will consider if there is any significant difference in the performance of the system when it is trained with PD data compared to being trained with healthy control (HC) data. We tailored our training procedure to test our hypothesis by performing variations of cross-validation. Apart from training our models on all HC subjects (ALL-HC), we also perform four different kinds of cross-validation: 1) We train our models on one PD subject (LOO-PD), 2) We train our models on one HC subject (LOO-HC), 3) We take one HC subject and use only roughly four minutes worth of data to train our models (4m-HC), 4) We take one PD subject and use only roughly four minutes worth of data to train our models (4m-PD). For all of our experiments, we test our trained models on all PD subjects (excluding the one used as training data for LOO-PD and 4m-PD). For room-level localisation accuracy, we use precision and weighted F1-score, all averaged and standard deviated across the test folds. 

To showcase the importance of in-home gait speed features in differentiating the medication state of a person with PD, we first compare how accurate the `Room-to-room Transition' duration produced by each network is to the ground truth (i.e. annotated location). We hypothesise that the more accurate the transition is compared to the ground truth, the better mobility features are for medication state classification. For the medication state classification, we then compare two different groups of features with two simple binary classifiers: 1) the baseline demographic features (see Section \ref{sec:cohort}), and 2) the normalised in-home gait speed features. The metric we use for ON / OFF medication state evaluation is the weighted F1-Score and AUROC which are averaged and standard deviated across the test folds.

\begin{table}[t!]
	\centering
        \setlength\tabcolsep{1.5pt}
        \caption{Room-to-room transition accuracy (in seconds) of all models compared to the ground truth. Standard deviation is shown in $(.)$, the best performer is bold, while the second best is italicized. A model that fails to capture a transition between particular rooms within a period that has the ground truth is assigned `N/A' score.}
	\begin{tabular}{ccccc}
		\toprule[1pt]
		\textbf{Data} & \textbf{Models} & \textbf{Kitch-Livin} & \textbf{Kitch-Dinin} & \textbf{Dinin-Livin} \\ 
		\midrule[0.5pt]
		\multicolumn{2}{c}{\textbf{Ground Truth}} & 18.71(18.52) & 14.65(6.03) & 10.64(11.99) \\
		\midrule[0.5pt]
		\multirow{6}{*}{\textbf{ALL-HC}} & RF & \textit{16.18(12.08)} & \textbf{14.58(10.22)} & 10.19(9.46) \\
		& TENER & 15.58(8.75) & 16.30(12.94) & 12.01(13.01) \\
		& Alt DTML &  15.27(7.51) & 13.40(6.43) & \textit{10.84(10.81)} \\
		& MDCSA & \textbf{17.70(16.17)} & \textit{14.94(9.71)} & \textbf{10.76(9.59)} \\
		\midrule[0.5pt]
		\multirow{6}{*}{\textbf{LOO-HC}} & RF & \textit{17.52(16.97)} & 11.93(10.08) & 9.23(13.69) \\
		& TENER & 14.62(16.37) & 9.58(9.16) & 7.21(10.61) \\
		& Alt DTML & 16.30(17.78) & \textit{14.01(8.08)} & \textbf{10.37(12.44)} \\
		& MDCSA & \textbf{17.70(17.42)} & \textbf{14.34(9.48)} & \textit{11.07(13.60)} \\
		\midrule[0.5pt]
		\multirow{6}{*}{\textbf{LOO-PD}} & RF & 14.49(15.28) & 11.67(11.68) & 8.65(13.06) \\
		& TENER & 13.42(14.88) & 10.87(10.37) & 6.95(10.28) \\
		& Alt DTML & \textbf{16.98(15.15)} & \textit{15.26(8.85)} & \textit{9.99(13.03)} \\
		& MDCSA & \textit{16.42(14.04)} & \textbf{14.48(9.81)} & \textbf{10.77(14.18)} \\
		\midrule[0.5pt]
		\multirow{6}{*}{\textbf{4m-HC}} & RF & 14.22(18.03) & 11.38(15.46) & 13.43(18.87) \\
		& TENER & 10.75(15.67) & 8.59(14.39) & N/A \\
		& Alt DTML & \textit{16.89(18.07)} & \textbf{14.68(13.57)} & \textit{9.31(15.70)} \\
		& MDCSA & \textbf{18.15(19.12)} & \textit{15.32(14.93)} & \textbf{11.89(17.55)} \\
		\midrule[0.5pt]
		\multirow{6}{*}{\textbf{4m-PD}} & RF & 11.52(16.07) & 8.73(12.90) & N/A \\
		& TENER & 8.75(14.89) & N/A & N/A \\
		& Alt DTML & \textit{14.75(13.79)} & \textit{13.47(17.66)} & N/A \\
		& MDCSA & \textbf{17.96(19.17)} & \textbf{14.74(10.83)} & \textbf{10.16(14.03)} \\
		\bottomrule[1pt]
	\end{tabular}
	\label{tab:transition_results}
\end{table}

\begin{table*}[t!]
    \centering
    \setlength\tabcolsep{6pt}
    \caption{PD participant room transition duration with ON and OFF medications comparison using Wilcoxon signed rank tests.}
    \begin{tabular}{cc|cc|ccc}
        \toprule[1pt]
        OFF transitions & Mean transition duration & ON transitions & Mean transition duration & W & z & p \\
        \midrule[0.5pt]
        Kitchen-Living OFF & 17.2 sec & Kitchen-Living ON & 14.0 sec & 75.0 & 2.824 & 0.001\\
        \midrule[0.5pt]
        Dining-Kitchen OFF & 12.9 sec & Dining-Kitchen ON & 9.2 sec & 76.0 & 2.903 & < .001\\
        \midrule[0.5pt]
        Dining-Living OFF & 10.4 sec & Dining-Living ON & 9.0 sec & 64.0 & 1.961 & 0.026\\
        \bottomrule[1pt]
    \end{tabular}
    \label{tab: clinical_correlation}
\end{table*}

\subsection{Experimental Results}

\textbf{Room-level Accuracy.} The first part of Table \ref{tab:benchmark_results} compares the performance of MCDSA network and other approaches for room-level classification. For the room-level classification, MDCSA network outperforms other networks and RF with a minimum improvement of 1.3\% for the F1-score over the second-best network (i.e. Alt DTML) in each cross-validation type with the exception of the ALL-HC validation. The improvement is more significant on the 4m-HC and 4m-PD validations, when the training data are limited, with an average improvement at almost 9\% for the F1-score over the Alt DTML.  

The LOO-HC and LOO-PD validation show that a model that has the ability to capture the temporal dynamics across time steps (e.g. TENER and DTML) will perform better than a standard baseline technique such as a Random Forest. TENER and DTML perform better in those two validations due to their ability to capture asynchronous relation across modalities. However, when the training data becomes limited as in 4m-HC and 4m-PD validations, having extra capabilities is necessary to further extract temporal information and correlations. Due to being a vanilla transformer requiring considerable amount of training data, TENER performs worst in these two validations. DTML performs quite well due to its ability to capture local context via LSTM for each modality. However, in general, DTML's performance suffers in both the LOO-PD and 4m-PD validations as the accelerometer data (and modality) may be erratic due to PD and should be excluded at times from contributing to room classification. MDCSA network has all the capabilities that DTML has with an improvement in suppressing accelerometer modality when needed via GRN layer embedded in DCSA. Suppressing the noisy modality seems to have a strong impact in maintaining the performance of the network when the training data is limited. This is validated by how Alt DTML (i.e. DTML added with GRN and CRF layers) outperforms the standard DTML by an average of 2.2\% for the F1-score in in 4m-HC and 4m-PD validations. It is further confirmed by MDCSA$_{1,4,7}$ 4APS against MDCSA$_{1,4,7}$ 4APS RSSI with the latter model, which does not include the accelerometer data, outperforming the former for the F1-score by an average of 1.6\% in the last three cross validations. It is worth pointing out that the MDCSA$_{1,4,7}$ 4APS RSSI model performed the best in the 4m-PD validation. However, the omission of accelerometer data affects the model in differentiating rooms that are more likely to have active movement (i.e. hall) than the rooms that are not (i.e. living room). It can be seen from Table \ref{tab:hallway_result} that the MDCSA$_{1,4,7}$ 4APS RSSI model has low performance in predicting hallway compared to the full model of MDCSA$_{1,4,7}$. As a consequence, the MDCSA$_{1,4,7}$ 4APS RSSI model cannot produce in-home gait speed features as competent as the ones produced by the MDCSA$_{1,4,7}$. 

\textbf{Room-to-room Transition and Medication Accuracy.} We hypothesise that during their OFF medication state, the deterioration in mobility of a person with PD is exhibited by how they transition between rooms. To test this hypothesis, a Wilcoxon signed rank test was used on the annotated data from PD participants undertaking each of the three individual transitions between rooms whilst ON (taking) and OFF (withholding) medications to assess whether the mean transition duration ON medications was statistically significantly shorter than the mean transition duration for the same transition OFF medications for all transitions studied (see Table \ref{tab: clinical_correlation}). From this result, we argue that the mean transition duration obtained by each model from Table \ref{tab:benchmark_results} that is close to the ground truth can capture what the ground truth captures. As mentioned in Section \ref{sec:cohort}, this transition duration for each model is generated by the model continuously performing room-level localisation focusing on the time a person is predicted to spend in a hallway between rooms. We show, in Table \ref{tab:transition_results}, that the mean transition duration for all transitions studied produced by MDCSA$_{1,4,7}$ model is the closest to the ground truth improving over the second best by around 1.25 seconds across all hall transitions and validations. 

The second part of the Table \ref{tab:benchmark_results} shows the performance of all our networks for medication state classification. The demographic features can be used as a baseline for each type of validation. The MDCSA network, with the exception of the ALL-HC validation, outperforms any other network by a significant margin for the AUROC score. By using in-home gait speed features produced by MDCSA network, a minimum of 15\% improvement over the baseline demographic features can be obtained with the biggest gained obtained in the 4m-PD validation data. In 4m-PD validation data, RF, TENER, and DTML could not manage to provide any prediction due to their inability to capture (partly) hall transitions. Furthermore, TENER has shown its inability to provide any medication state prediction from the 4m-HC data validations. It can be validated by Table \ref{tab:transition_results} when the TENER failed to capture any transitions between dining room and living room across all periods that have ground truths. MDCSA networks are able to provide medication state prediction and maintain its performance across all cross-validations thanks to the addition of Eq. \ref{eq:bce} in the loss function. 

\section{Conclusion}
\label{sec:conclusion}

We have presented the MDCSA model, a new deep learning approach for indoor localisation utilising RSSI and wrist-worn accelerometer data. The evaluation on our unique real-world free-living pilot dataset, which includes subjects with and without PD, shows that MDCSA achieves the state-of-the-art accuracy for indoor localisation. The availability of accelerometer data does indeed enrich the RSSI features which, in turn, improves the accuracy of the indoor localisation. 

In naturalistic settings, in-home mobility can be measured through the use of indoor localisation models. We have shown, using room transition duration results, that our PD cohort take longer on average to perform a room transition when they withhold medications.  With accurate in-home gait speed features, a classifier model can then differentiate accurately if a person with PD is in an ON or OFF medication state. Such changes show the promise of these localisation outputs to detect the dopamine-related gait fluctuations in PD that impact on patients' quality of life \cite{Marras2004} and are important in clinical decision making \cite{Clarke2009}.  We have also demonstrated that our indoor localisation system provides precise in-home gait speed features in PD with a minimal average offset to the ground truth. The network also outperforms other models in the production of in-home gait speed features which is used to differentiate the medication state of a person with PD. 


\textbf{Limitations and future research.} One limitation of this study is the relatively small sample size (which was planned as this is an exploratory pilot study). We believe our sample size is ample to show proof of concept. This is also the first such work with unobtrusive ground truth validation from embedded cameras. Future work should validate our approach further on a large cohort of people with PD and consider stratifying for sub-groups within PD (e.g. akinetic-rigid or tremor-dominant phenotypes), which would also increase the generalisability of the results to the wider population. Future work in this matter could also include the construction of a semi-synthetic dataset based on collected data to facilitate a parallel and large-scale evaluation. 



This smart home’s layout and parameters remain constant for all the participants, and we acknowledge that transfer of this deep learning model to other varied home settings may introduce variations in localisation accuracy. For future ecological validation and based on our current results, we anticipate the need for pre-training (e.g. a brief walkaround which is labelled) for each home, and also suggest that some small amount of ground-truth data will need to be collected (e.g. researcher prompting of study participants to undertake scripted activities such as moving from room to room) to fully validate the performance of our approach in other settings. 

Accurate room localisation using these data modalities has a wide range of potential applications within healthcare. This could include tracking of gait speed during rehabilitation from orthopaedic surgery, monitoring wandering behaviour in dementia or triggering an alert for a possible fall (and long lie on the floor) if someone is in one room for an unusual length of time. Furthermore, accurate room use and room-to-room transfer statistics could be used in occupational settings, e.g. to check factory worker location. 

\begin{acks}
We are very grateful to the study participants for giving so much time and effort to this research. We acknowledge the local Movement Disorders Health Integration Team (Patient and Public Involvement Group) for their assistance at each study design step. This work was supported by the SPHERE Next Steps Project funded by the UK Engineering and Physical Sciences Research Council (EPSRC), [Grant EP/R005273/1]; the Elizabeth Blackwell Institute for Health Research, University of Bristol and the Wellcome Trust Institutional Strategic Support Fund [grant code: 204813/Z/16/Z]; by Cure Parkinson’s Trust [grant code AW021]; and by IXICO [grant code R101507-101]. Dr and Mrs de Pass made a charitable donation to the University of Bristol through the Development and Alumni Relations Office to pay the salary of CM, the PhD student and Clinical Research Fellow.
\end{acks}

\bibliographystyle{ACM-Reference-Format}
\bibliography{cas-refs}

\newpage
\appendix

\section{Statistical Significance Test}

\begin{figure}[h]
    \centering
    \includegraphics[width=0.5\textwidth]{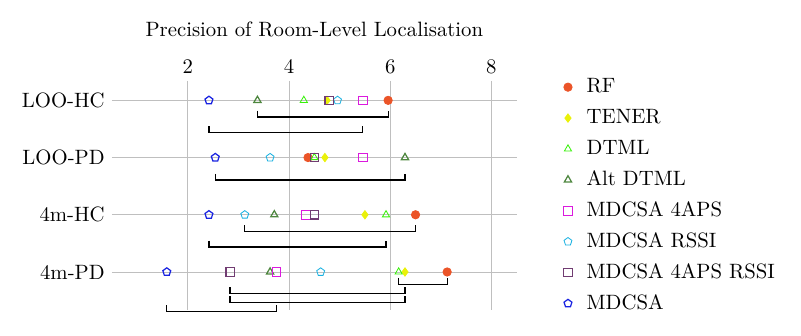}
    \caption{Critical difference diagram for the precision of room-level localisation showing the pairwise statistical comparison of all localisation models across different cross-validation techniques.} 
    \label{fig:f1_score_CDD}
\end{figure}

\begin{figure}[h]
    \centering
    \includegraphics[width=0.5\textwidth]{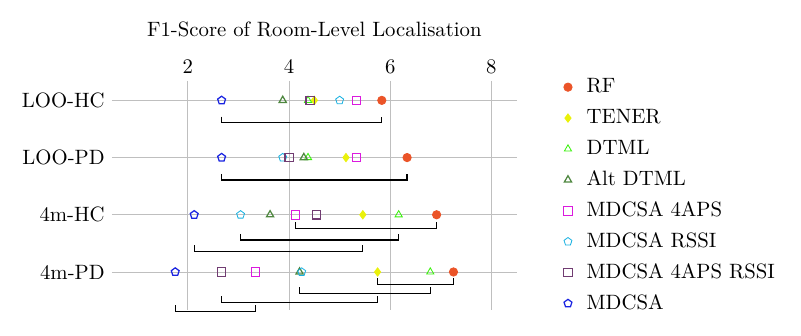}
    \caption{Critical difference diagram for the F1-score of room-level localisation showing the pairwise statistical comparison of all localisation models across different cross-validation techniques.} 
    \label{fig:precision_CDD}
\end{figure}

It could be argued that all the localisation models compared in Table \ref{tab:benchmark_results} might not be statistically different due to fairly high standard deviation across all types of cross-validations which is caused by relatively small number of participants. In order to compare multiple models over cross-validation sets and show statistical significance of our proposed model, we perform the Friedman test to first reject the null hypothesis \cite{demsar2006statistical}. We then performed a pairwise statistical comparison: the Wilcoxon signed-rank test with Holm’s alpha correction ($\alpha = 5\%$). Finally, we used a critical difference diagram \cite{benavoli2016should} to visualize the results of these statistical tests projected onto the average rank axis, with a thick horizontal line showing a clique of localisation mdoels that are not significantly different (see Figure \ref{fig:f1_score_CDD} and \ref{fig:precision_CDD}).

\end{document}